\documentclass[prl,aps,reprint,nobibnotes,twocolumn]{revtex4-1}

\usepackage{amsmath,amssymb,graphicx,color}
\usepackage{soul} %\st \hl
\usepackage{braket}

\soulregister\ref{7}
\soulregister\eqref{7}
\soulregister\cite{7}
\soulregister\onlinecite{7}

\begin{document}

\title{Generating optical vortex beams by momentum-space polarization vortices centered at bound states in the continuum}

\author{Bo Wang\textsuperscript{1}}
\thanks{These authors contributed equally to this work.}
\author{Wenzhe Liu\textsuperscript{1,*}}
\email{wzliu15@fudan.edu.cn}
\author{Maoxiong Zhao\textsuperscript{1,*}}
\author{Jiajun Wang\textsuperscript{1}}
\author{Yiwen Zhang\textsuperscript{1}}
\author{Ang Chen\textsuperscript{1}}
\author{Fang Guan\textsuperscript{1}}
\author{Xiaohan Liu\textsuperscript{1,2}}
\author{Lei Shi\textsuperscript{1,2}}
\email{lshi@fudan.edu.cn}
\author{Jian Zi\textsuperscript{1,2}}
\email{jzi@fudan.edu.cn}
\affiliation{$^{1}$ State Key Laboratory of Surface Physics, Key Laboratory of Micro- and Nano-Photonic Structures (Ministry of Education) and Department of Physics, Fudan University, Shanghai 200433, China}
\affiliation{$^{2}$ Collaborative Innovation Center of Advanced Microstructures, Nanjing University, Nanjing 210093, China}

\begin{abstract}
An optical vortex (OV) is a beam with spiral wave front and screw phase dislocation. This kind of beams is attracting rising interest in various fields. Here we theoretically proposed and experimentally realized a novel but easy approach to generate optical vortices. We leverage the inherent topological vortex structures of polarization around bound states in the continuum (BIC) in the momentum space of two dimensional periodic structures, e.g. photonic crystal slabs, to induce Pancharatnam-Berry phases to the beams. This new class of OV generators operates in the momentum space, meaning that there is no real-space center of structure. Thus, not only the fabrication but also the practical alignment would be greatly simplified. Any even order of OV, which is actually a quasi-non-diffractive high-order quasi-Bessel beam, at any desired working wavelength could be achieved in principle. The proposed approach expands the application of bound states in the continuum and topological photonics.
\end{abstract}

\maketitle

An optical vortex (OV) is a light beam with spiral phase front and a zero-intensity point at the beam center. OVs are proved to be carrying orbital angular momentum (OAM) \cite{allen1992orbital, bliokh2006geometrical, franke2008advances, dennis2009singular, bliokh2010angular, yao2011orbital}, which is a new degree of freedom of light and has greatly broadened the fields such as optical microscopy \cite{furhapter2005spiral}, optical micromanipulation \cite{t2000three, o2002intrinsic, grier2003revolution, curtis2002dynamic, curtis2003structure, ng2010theory, padgett2011tweezers}, optical communications \cite{gibson2004free, walker2012trans, willner2015optical} and quantum information processing \cite{mair2001entanglement, vaziri2002experimental, molina2007twisted}. As a result, generating OVs becomes a very hot topic in different wavelength ranges like visible light, microwaves and radio waves \cite{karimi2014generating, zhang2018phase, tamburini2011experimental, miao2016orbital}. In low frequency range, spiral phase plates \cite{beijersbergen1994helical} and phased antenna arrays \cite{bai2014experimental} are the most commonly used approaches to generate OVs. However, both those two kinds of structures can hardly be made into compact and integrable devices when the working wavelength becomes smaller approaching the terahertz or visible range, due to the limitations on device thickness and feed network arrangements. Recently, with rapid development of metasurfaces \cite{huang2012dispersionless, kildishev2013planar, yu2014flat, liu2014broadband}, planar OV generators in micrometer or nanometer scale are shown possible, still requiring intricate designing of the individual units, complexity in fabrication to introduce helical phases and tough alignment in practice.

In this paper, we propose that, instead of artificially arranging resonators with winding configurations in the real space, we can induce Pancharatnam-Berry (PB) phases \cite{berry1987adiabatic} to beams by taking the advantage of winding topologies of resonances which naturally exist in the momentum space near bound states in the continuum (BIC) \cite{zhen2014topological, hsu2016bound, zhang2018observation, doeleman2018experimental, chen2019observing, chen2019singularities}. We realize such kind of OV generators with photonic crystal (PhC) slabs \cite{fan2002analysis} which are very simple in structure. Operating in the momentum space, the proposed OV generators have no center of structure to be aligned at the incident beam center. The generated OVs are proved to have ring-like profiles in the momentum space, thus they are quasi-Bessel beams which have quasi-diffraction-free behavior \cite{vaity2015perfect} (their momentum space profiles are rings with finite peak widths rather than $\delta$-function rings as perfect Bessel beams). By changing only the symmetry and the scale of the unit cells, different orders of OVs at visible and near-infrared working wavelengths are experimentally achieved.

\begin{figure}[t]
\centering
% Requires \usepackage{graphicx}
\includegraphics[scale=1]{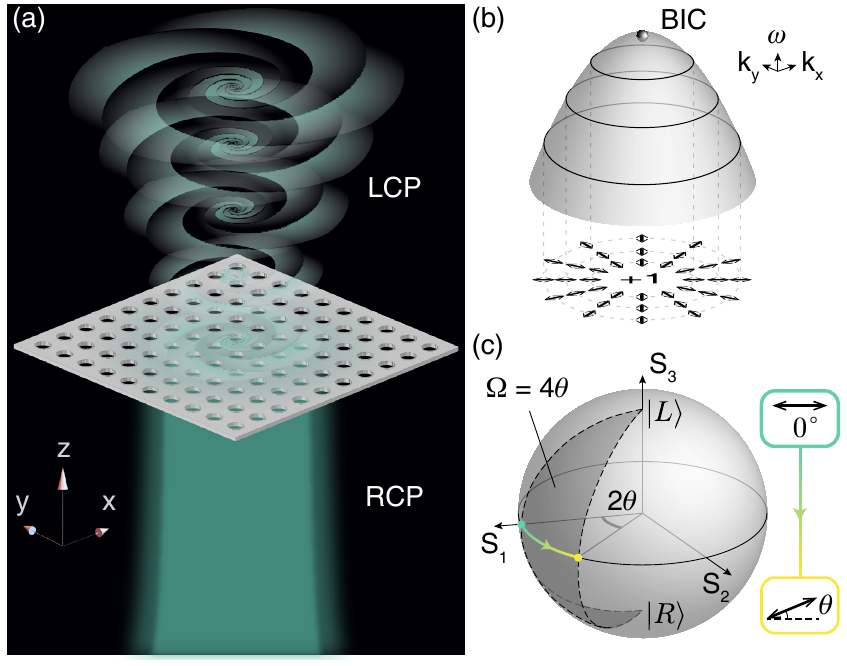}
\caption{\textbf{Concept of the proposed optical vortex generation method.} (a) A schematic view of proposed optical vortex (OV) generating approach. (b) A schematic view of a parabolic band of a photonic crystal (PhC) slab, which has a bound state in the continuum (BIC) in the center. The vortex structures formed by nearly-linear guided resonances at different frequencies close to the BIC frequency are projected onto the momentum space plane. (c) The working principle of the OV generator. Different linear states of polarization (SOPs) correspond to different azimuthal positions on the equator of the Poincar\'e sphere, which will determine the trajectory of the SOPs and result in different geometric phase factors.}
\label{fig:1}
\end{figure}

As plotted in Fig. \ref{fig:1}(a), a 4-fold rotational-symmetric photonic crystal slab can be viewed as one example of our proposed OV generator. Unlike various metasurface-based OV generators, it seems counterintuitive that, there are neither space-variant resonators nor winding topologies in the system to induce the extra spiral phase factors. Actually, there ARE underlying winding topologies in these radiative periodic systems. They exist in the momentum space. Recently, vortex structures of resonances are theoretically studied \cite{zhen2014topological, chen2019singularities, sadrieva2019multipolar, chen2019line} and experimentally observed \cite{zhang2018observation, doeleman2018experimental, chen2019observing} in the momentum space of 2d periodic structures such as PhC slabs, two dimensional plasmonic crystals and gratings. It is of great importance that those vortex topologies are tightly related to the fascinating optical phenomenon, BIC \cite{bulgakov2008bound, hsu2013observation, zhen2014topological, bulgakov2014bloch, yang2014analytical, hsu2016bound, gomis2017anisotropy, guo2017topologically, bulgakov2017bound, zhang2018observation, doeleman2018experimental, song2018cherenkov, dai2018topologically, he2018toroidal, jin2018topologically, koshelev2018asymmetric, koshelev2018meta, guo2019arbitrary, chen2019singularities, cerjan2019bound, koshelev2019nonradiating, sadrieva2019multipolar, chen2019line, chen2019observing, cerjan2019bound}, and they are believed to result from topological property of the system. For example, let us consider a PhC slab with rotational symmetry higher than 2-fold [4-fold in Fig. \ref{fig:1}(a)]. In such a system, any singlet at the $\mathrm{\Gamma}$ point must be a BIC. In the vicinity of the BIC, the resonant guided modes are of high quality factors. Moreover, the states of polarization (SOPs) of far-field radiation from these guided resonances, which are almost linear, are momentum-space-variant and forced by symmetry to form vortex topologies, shown as Fig. \ref{fig:1}(b). In other words, the whole PhC slab will behave as different polarized resonators with different incident wave vector $k$ near a at-$\mathrm{\Gamma}$ BIC. These ``momentum-space resonators" certainly could play the role that the real-space winding geometry played in the previous works.

\begin{figure}[b]
\centering
% Requires \usepackage{graphicx}
\includegraphics[scale=1]{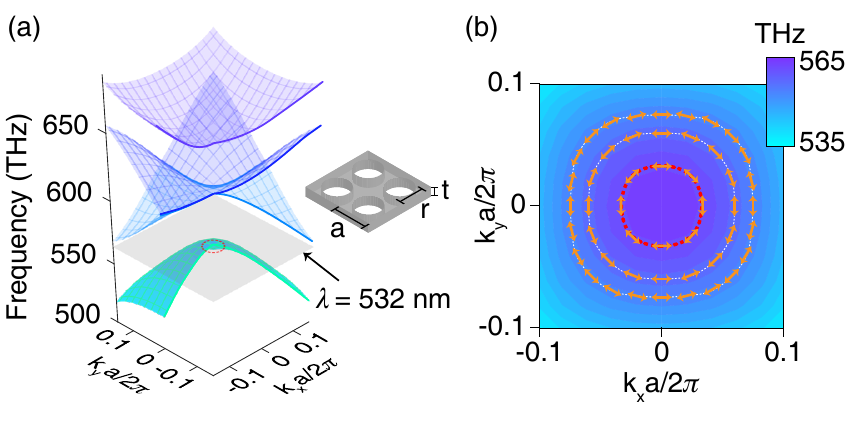}
\caption{\textbf{Simulated band structure and polarization distribution of the designed $C_{4v}$ structure.} (a) Simulated band structure of the designed sample near the $\mathrm{\Gamma}$ point. The band we focus on (TE-like 2) is colored with opaque dark turquoise, while the other TE-like bands are translucent. The band surfaces are sliced in $\mathrm{\Gamma-X}$ and $\mathrm{\Gamma-M}$ direction. The working wavelength (532 nm) is pointed out with a translucent dark plane. Inset: the designed structure. (b) The iso-frequency contours (dotted loops) with the SOPs (orange double-sided arrows) marked on them. The contours correspond to 532 (the working wavelength, marked red), 535, 538 nm from inside to outside. The background is the flattened band surface of TE-like 2, of which the different colors correspond to different frequencies.}
\label{fig:2}
\end{figure}

\begin{figure*}[t]
\centering
% Requires \usepackage{graphicx}
\includegraphics[scale=1]{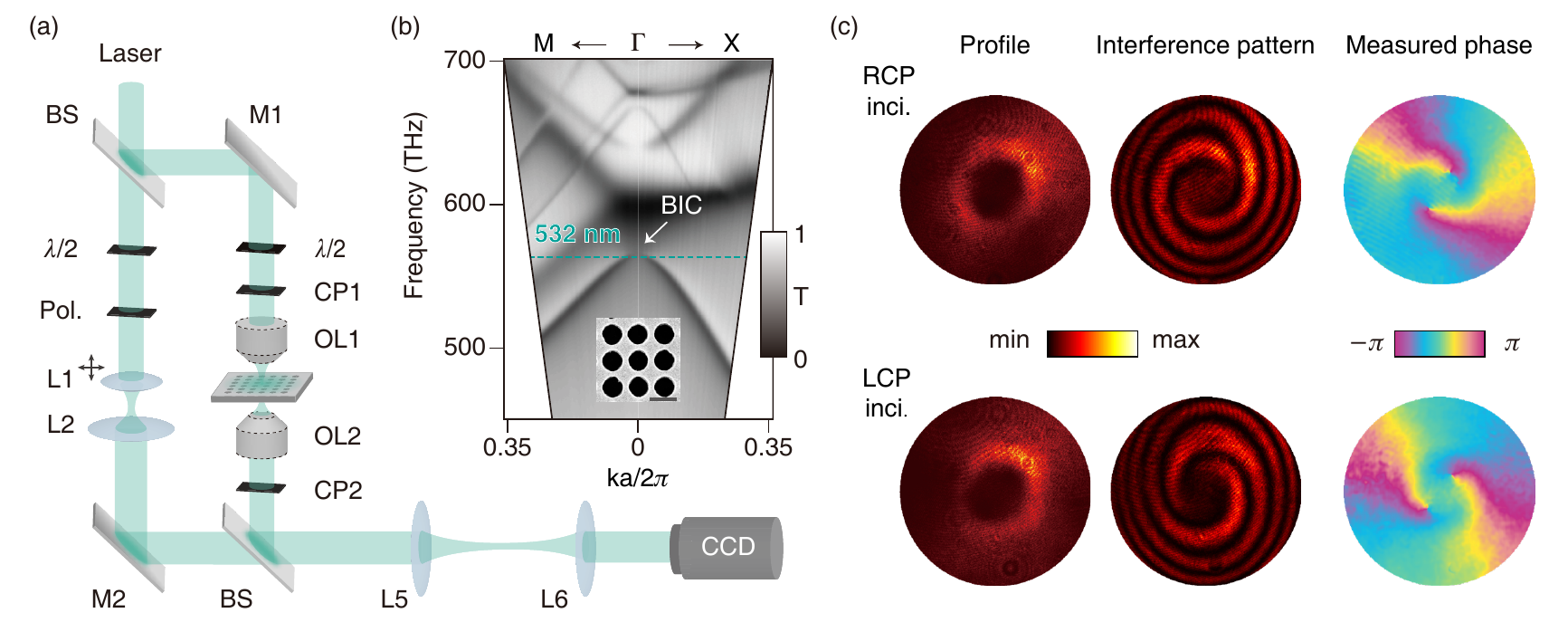}
\caption{\textbf{Experimental setup and the measured results of the fabricated 4-fold symmetric sample.} (a) The experimental setup. The lens L1 in the reference light path could be moved to modify the reference wave's direction and shape of wave front. (b) The dispersion of the sample measured as angle-resolved transmittance spectra. The working wavelength is marked with a green dashed line. The vanishing point of the transmittance signal on TE-like 2 corresponds to the central BIC. The white regions on the two sides are regions limited by the numerical aperture which cannot be measured. Inset: the scanning electron microscopy image of the sample (scale bar: 400 nm). (c) The measured beam profile, interference pattern and phase distribution of the transmitted cross-polarized beams. The angular range of the plots are about $13^\circ$. Note that L1 is moved a bit making the reference beam spherical to show the vortex-shaped interference pattern more obviously.}
\label{fig:3}
\end{figure*}

Choosing a working frequency close to the at-$\Gamma$ BIC and regarding the guided resonances of which the SOPs are almost linear as wave plates [oriented in the direction $\theta(k_{\parallel})$ to the $x$ axis], the transmitted field of a certain $k_{\parallel}$ incidence can be formulated as \cite{bomzon2002space}
\begin{equation*}
  \begin{aligned}
    \left| E_{\mathrm{out}} \right> = \\
    &
    \frac{1}{2}
    [t_x(k_{\parallel}) + t_y(k_{\parallel})]
    \left| E_{\mathrm{in}} \right>
    + \\
    &
    \frac{1}{2}
    [t_x(k_{\parallel}) - t_y(k_{\parallel})]e^{-2i\theta(k_{\parallel})}
    \braket{E_{\mathrm{in}}|R}
    \ket{L}
    + \\
    &
    \frac{1}{2}
    [t_x(k_{\parallel}) - t_y(k_{\parallel})]e^{2i\theta(k_{\parallel})}
    \braket{E_{\mathrm{in}}|L}
    \ket{R}
  \end{aligned}
\end{equation*}
on a helical basis. Here $t_x, t_y$ are the transmittance coefficients of the $k_{\parallel}$ guided resonance with the polarization parallel and perpendicular to the efficient fast axis. $\ket{E_{\mathrm{in}}}, \ket{E_{\mathrm{out}}}$ are the Jones vectors of the incident and transmitted light, while $\ket{L}, \ket{R}$ denote the left- and right-handed circularly polarized (LCP \& RCP) unit vectors $(0,1)^\mathrm{T}$ \& $(1,0)^\mathrm{T}$. From the formula it is clear that, if we choose the incident light to be circularly polarized, the transmitted light would be composed of a trivial part with the same polarization of the incidence, along with another non-trivial cross-polarized part converted by the resonant process. This part of light would gain a geometric phase factor, i.e. Pancharatnam-Berry (PB) phase \cite{berry1987adiabatic}, depending on the orientation of the SOP of the guided resonance, which can also be understood more phenomenologically by introducing the Poincar\'e sphere picture [See Fig. \ref{fig:1}(c)]. The PB phase equals to half the solid angle enclosed by the trajectory of SOPs on the Poincar\'e sphere. With the starting point and the end pinned at the opposed poles, the trajectory (also the PB phase) would vary with the intermediate point on the equator which correspond to the orientation of the resonance in the $k$-space. As a result, when we normally shine a right-handed circularly-polarized (RCP) and slightly divergent beam at the corresponding working wavelength onto the PhC slab, the different $k$ components of the beam would interact with different $k_{\parallel}$ resonances, then the transmitted left-handed circularly-polarized (LCP) beam would gain the desired spiral phase front of which the topological charge is
\begin{equation*}
  l = -2 \times q
\end{equation*}
($q$ here is the polarization charge of the BIC). The detailed proof could be found in the Supplemental Material (SM) \cite{sm}. Furthermore, if we choose the guided resonances to be on a parabolic (or conical) band, the amplitude distribution will be a circular ring (the same shape as iso-frequency contours) in the $k$-space, indicating that the generated OV is actually a high-order quasi-Bessel beam. The quasi-non-diffractive nature of the quasi-Bessel beam is very meaningful. By applying a lens, the transmitted beam can be Fourier transformed into a quasi-Laguerre-Gaussian one.

Note that, we here only present the theory in transmitting mode. However, the proposed approach works fine in reflecting mode with the theory almost the same. Our OV generating approach basing on Bloch modes only requires symmetry and periodicity. Adding a substrate even a mirror or changing the basing material to even metal won't matter as long as the Bloch modes exist. The wide choice of materials and the simple structure would dramatically simplify the designing and fabrication for practical uses. Accounting that our proposed OV generators work in the momentum space, no real-space alignment according to the optical axis is needed in application. In addition, periodic structures like PhC slabs can mostly be scaled up or down arbitrarily to work in different wavelength ranges like microwaves.

To experimentally realize the proposed approach, we firstly designed a PhC slab working at 532 nm, which is in the visible range. We obtained its band structure and SOP distribution in the momentum space by simulations in order to assure that our proposition would work. Fig. \ref{fig:2} shows the structure along with the simulated results. The structure shown as the inset of Fig. \ref{fig:2}(a) is a periodically etched freestanding silicon nitride (Si$_3$N$_4$, refractive index $n \approx 2.02$) slab, of which the thickness $t$ is 120 nm. The lattice is square and $C_{4}$ symmetric, and the periodicity $a$ is 380 nm. The etched holes are circular with their radius $r$ equal to 140 nm. The calculated band structure is illustrated in Fig. \ref{fig:2}(a), and SOPs at three different wavelengths on band TE-like 2 are shown in Fig. \ref{fig:2}(b). One can clearly see that this band we focus on is a paraboloid, making the iso-frequency contours near the $\mathrm{\Gamma}$ point circularly shaped as we wish. More importantly, the SOPs on the iso-frequency contours show predicted winding behaviors corresponding to the central BIC, which allow us to induce geometric phases. The total winding angle of the SOPs is $2\pi$ along a counterclockwise loop around the polarization singularity, i.e. the topological charge of the BIC is $q = +1$. The SOPs are all close to linear polarization, which verifies our approximation considering the guided resonances as wave plates. It needs to be emphasized again that, although our structure has extra symmetries such as mirror symmetries and is made of a dielectric material, the only necessary condition to design such kind of OV generators is a $i$-fold ($i > 2$) rotational symmetry, which sustains the existence of parabolic bands and BICs on them.

We then fabricated the designed prototype of our proposed OV generator. A freestanding Si$_3$N$_4$ layer window on a silicon substrate is periodically etched applying reactive-ion etching technique. The parameters are the same as designed, with the total number of unit cells being $260 \times 260$. A scanning electron microscopy image is given in the inset of Fig. \ref{fig:3}(b). We built a home-made Fourier-optics-based spectroscopy system, which can operate in three modes: a spectrometer mode, an imaging mode and an interferometer mode. With this powerful system, we not only manage to measure the profile and the phase of the transmitted beam, but also can measure the angle-resolved spectra of the sample. The schematic of the experimental setup especially for the interferometer mode is shown in Fig. \ref{fig:3}(a).

First of all, the system operates in the spectrometer mode of which the details could be found in Ref. \cite{zhang2018observation}. In this mode, the angle-resolved transmittance spectra along each azimuthal direction of the sample can be measured with a single shot. From the measured spectra plotted as Fig. \ref{fig:3}(b), the dispersion of the fabricated prototype is exhibited explicitly. We find the results agree with our simulated band structure well, and the at-$\mathrm{\Gamma}$ state is indeed a BIC which cannot be excited and makes the Fano-resonance feature of the spectrum disappear.

Subsequently, we switch the system to the imaging mode, in which no reference beam is applied. Passing through the incident circular polarizer (CP1) and focused by an objective lens (OL1), the incident beam is convergent and circularly polarized. The beam will excite the $k$-variant guided resonances close to the BIC, generating a cross-polarized counterpart with a phase vortex. Then, the transmitted beam is filtered by the orthogonally polarized CP2 and fourier-transformed into far field by another objective lens (OL2). We finally obtain the far-field profiles of the cross-polarized beams from the CCD, plotted as the first column of Fig. \ref{fig:3}(c). One can see the beam profiles in far field are donut-shaped, confirming that they are quasi-Laguerre-Gaussian beams after Fourier transformation, and are originally quasi-Bessel beams.

In order to verify whether the filtered beams are OVs, we switched our system to the interferometer by introducing a reference beam. The reference beam is linear polarized and made a little divergent using a set of convex lenses (L1 \& L2) to show the interference pattern more clearly, as plotted in the second column of Fig. \ref{fig:3}(c). We find that there are two spiral arms in each interference pattern, proving the beams to be OVs with their topological charges $l = \mp2 = \mp2 \times q$. Exchanging the polarizer and analyzer, the spiral arms will change their directions. This corresponds to the fact that, the geometric phase vortex will change its sign when the start and end points of the trajectory on the Poincar\'e switch their locations. Also, we apply the method mentioned in Ref. \cite{zambon2019optically} based on Fourier component filtering to measure the phase distribution. The measured distributions are illustrated as the third column of Fig. \ref{fig:3}. The transmitted beam with RCP incidence apparently have a topological charge of -2, while the charge of the LCP one is +2, as predicted. The separation of the phase singularity is due to the defects in the sample and the imperfectness of the CPs. It is worth repeating that, because the working resonances are in the momentum space, the beam is not required to be focused on the center of the sample. As a proof, we have tried to move the sample in the experiment, finding the far-field beam profile unchanged. The result could be found in the SM \cite{sm}. We also simulated the propagating behavior of the generated beam, confirming that the beam could maintain diffraction free after a 7.5-micron (about 14 times the working wavelength) propagation, of which the results could also be found in the SM \cite{sm}.

\begin{figure}[b]
\centering
% Requires \usepackage{graphicx}
\includegraphics[scale=1]{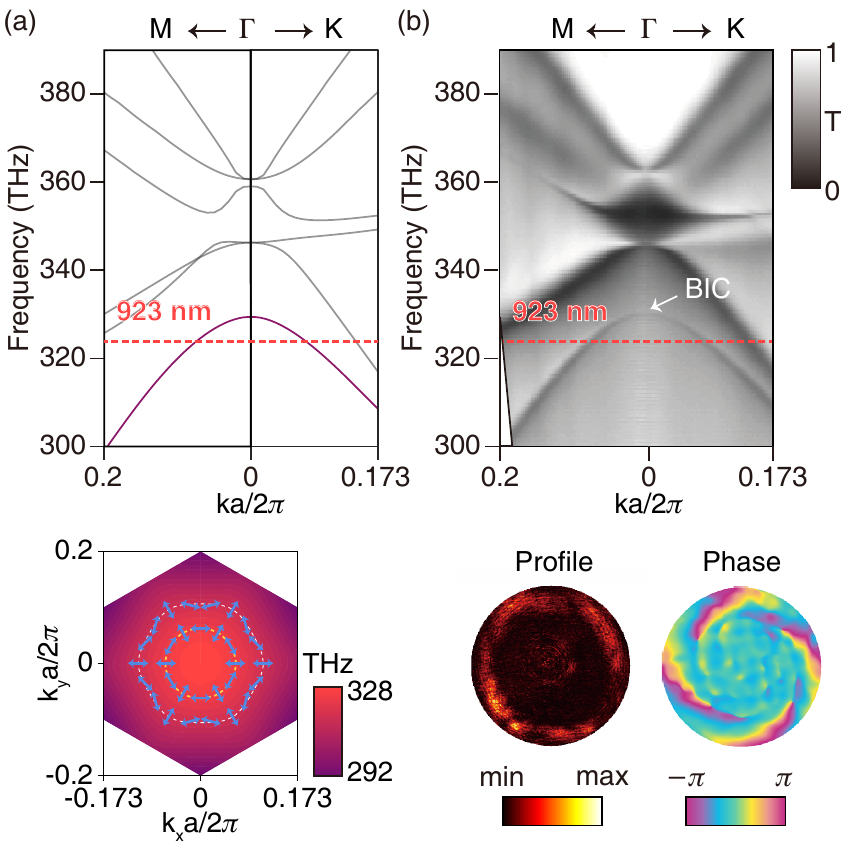}
\caption{\textbf{The measured results of the fabricated 6-fold symmetric sample generating a higher-order optical vortex.} (a) The simulated band structure of the $C_{6v}$ PhC slab and the polarization distribution on its band TE-like 2. In the band structure diagram, the band TE-like 2 is marked by purple and the working wavelength is marked by the red dashed line. In the polarization plot, the background is the flattened band surface of the studied band, while the dashed lines are the iso-frequency contours of 923 nm (marked orange) and 947 nm. (b) The measured band structure of the PhC slab in the form of transmittance spectra and the profile \& phase distribution of the generated beam with RCP incidence.}
\label{fig:4}
\end{figure}

Considering the simple working principle we proposed, OVs with higher topological charges can also be achieved by applying structures with higher symmetry. We here designed another sample with $C_{6v}$ symmetry, and the working wavelength is modified to be in the near-infrared range (923 nm). The thickness of the slab is modified to 100 nm, while the periodicity is 850 nm and the radius of holes is 265 nm. Similarly, we get the band structure and the polarization distribution of the $C_{6v}$ sample by simulation, shown in Fig. \ref{fig:4}(a). The applied band here is TE-like 2, of which the winding number of the central BIC is -2. Thus, the produced OVs shall be with topological charge $l = \mp2 \times q = \pm4$, which we are also able to measure experimentally. The experimental results are plotted as Fig. \ref{fig:4}(b). With RCP incidence, the generated OV has the charge of +4. For systems with higher symmetry, we can even change the topological order of the generated OV by switching the working wavelength. Experimental results switching charge of OV from $\pm4$ to $\mp2$ are shown in the SM \cite{sm}.

In conclusion, our proposed OV generating approach is of high feasibility and practicability with variable working wavelengths and orders of OVs. It also provides strategies to design BIC-vortex-lasers \cite{kodigala2017lasing,bahari2017integrated,ha2018directional}. Besides, no center alignment is needed as the approach operates in the momentum space. Since the at-$\mathrm{\Gamma}$ BIC phenomenon only relies on the periodicity and symmetry of the system, the presented approach of OV generating can also be expanded to any dielectric materials, semiconductors, and even metals. The generated OVs are naturally high order quasi-Bessel beams, and can be transformed to quasi-Laguerre-Gaussian beams using lenses. The proposed planar PhC slab structure is lightweight and highly integrable, which meets the increasing need of compact optical elements. Considering the fact that the winding behavior of SOPs around BICs comes from the topological property of the photonic bands, our work also introduce a new doorway to study and utilize topological photonics.

% Bibliography
\bibliography{OVgen_reference.new}

\section{Acknowledgements}

\begin{acknowledgments}
We thank Prof. Che Ting Chan, Dr. Haiwei Yin for helpful discussions. The work was supported by 973 Program and China National Key Basic Research Program (2015CB659400, 2016YFA0301100, 2016YFA0302000 and 2018YFA0306201) and National Science Foundation of China (11774063, 11727811 and 91750102). The research of L. S. was further supported by Science and Technology Commission of Shanghai Municipality (17ZR1442300, 17142200100).
\end{acknowledgments}

\section{Author contributions}
W. L., L. S. and J. Z. conceived the basic idea for this work. W. L. gave the theoretical explanation. B. W. and W. L. designed the structures, carried out the finite element method and the finite-difference time-domain method simulations, and analyzed the simulated and measured data. B. W. and J. W. performed the sample fabrications. B. W., M. Z. and J. W. performed the optical measurements. M. Z. and Y. Z. constructed the measurement system. L. S. and J. Z. supervised the research and the development of the manuscript. W. L. wrote the draft of the manuscript and all authors took part in the discussion, revision and approved the final copy of the manuscript.

\section{Data availability}
The data that support the findings of this study are available from the authors on reasonable request, see author contributions for specific data sets.

\clearpage

\section{Methods}

\textbf{Theoretical analysis.} Please see the Supplemental Materials for the derivations and discussions.

\textbf{Simulations.} The eigen-mode simulations and the polarization analysis were done using finite element method and finite-difference time-domain method. Periodic (Bloch) boundary conditions were applied in $x,y$ direction, while perfect matching layers were applied in $z$ direction.

\textbf{Sample fabrication.} The samples were fabricated basing on commercial silicon nitride windows, of which the silicon nitride layers are $100 \sim 120$ nanometers thick. The silicon nitride layers resided on center-windowed silicon substrates, whose thicknesses are about 200 microns. To fabricate a designed structure, the raw sample was firstly spin-coated with a layer of positive-tone electron-beam resist (PMMA950K A4, MicroChem). An additional layer of conductive polymer (AR-PC 5090.02) was also attached to avoid charging effects during electron-beam lithography (EBL). Then, a hole array mask pattern was defined onto the PMMA layer using EBL (ZEISS Sigma 300), followed by developing in a 1:3 mixture of methyl isobutyl ketone (MIBK) and isopropyl alcohol (IPA). The periodically EBL etched PMMA layer would act as a mask in the subsequent reactive-ion etching (RIE) process. Anisotropic etching of the periodic structure was achieved using a mixture of CHF$_3$ and O$_2$. After ensuring that the freestanding part of the silicon nitride layer had been etched through, the PMMA mask was eventually removed by RIE using O$_2$. The overall sizes of the fabricated samples are approximately 100 microns $\times$ 100 microns.

\textbf{Experimental technique.} The Fourier-optics-based momentum-space spectroscopy system has three operating modes: a spectrometer mode, an imaging mode, and an interferometer mode. The illustration of the system could be seen in Fig. \ref{fig:3}, while the schematic views of the other two modes could be found in Ref. \cite{zhang2018observation, chen2019observing}.

To switch the system to the spectrometer mode, the reference beam should be blocked, and the light source should be replaced with a broadband one. An spectrometer with a slit should be plugged in front of the charge coupled device (CCD) to resolve the frequencies, and the circular polarizers should be removed. Operating in this mode, the dispersion of the system could be obtained in the form of angle-resolved transmittance spectra.

Meanwhile, the imaging mode would enable us to obtain the beam profile. To use this mode, the light source should be a monochromatic laser, the spectrometer should be removed, and the polarizers should be plugged back in. Focused by an objective lens (OL1) and passing through the incident circular polarizer (CP1), the incident beam would be convergent and circularly polarized. With the analyzer (CP2), the cross-polarized part of the transmitted beam is filtered out and its profile would be captured by the CCD.

Thirdly, the interferometer mode of the system could give us the interference fringes and the phase distributions. The system could be switched to this mode by introducing the reference beam. When the lenses L1 and L2 were confocal, the reference beam should be plane wave like. Shifting the lens L1 away from confocal condition on-axis would make the reference beam spherical and the interference patterns would thus be more vivid vortex-like ones. Otherwise, if the lens L1 is shifted off-axis, the reference beam would gain an extra transverse wave vector, which would separate the phase distribution information from the zeroth-order Fourier component of the interference fringes to the $\pm1$st-order ones. By extracting the 1st- or the $-1$st-order Fourier component, the phase distributions of the beams could be retrieved.

\end{document}